\documentclass[pre,twocolumn,groupedaddress,showpacs,showkeys,amsmath,amssymb,floatfix]{revtex4}

\usepackage{graphicx}
\usepackage{dcolumn}
\usepackage{bm}

\usepackage{subfigure}

\graphicspath{{images/}{plots/}}

\newcommand{\brac}[1]{\ensuremath{\left(#1\right)}}

\newcommand{\change}[1]{#1}

\DeclareMathOperator{\Var}{Var}

\newcommand{\s}{\ensuremath{\sigma}}

\newcommand{\sC}{\ensuremath{\sigma_{c}}}
\newcommand{\bC}{\ensuremath{b}}

\newcommand{\st}{\ensuremath{\sigma^{\mathrm{\tau}}}}
\newcommand{\sac}{\ensuremath{\sigma_{c}^{\mathrm{cp}}}}
\newcommand{\bac}{\ensuremath{{b^{\mathrm{cp}}}}}

\newcommand{\sacV}{\ensuremath{1.07(5)}}
\newcommand{\bacV}{\ensuremath{0.43(3)}}

\newcommand{\sdc}{\ensuremath{\sigma_{c}^{\mathrm{lp}}}}
\newcommand{\bdc}{\ensuremath{b^{\mathrm{lp}}}}

\newcommand{\sdcV}{\ensuremath{0.51(4)}}
\newcommand{\bdcV}{\ensuremath{0.29(6)}}

\newcommand{\sfc}{\ensuremath{\sigma_{c}^{\mathrm{fb}}}}
\newcommand{\bfc}{\ensuremath{b^{\mathrm{fb}}}}

\newcommand{\sfcV}{\ensuremath{1.47(8)}}
\newcommand{\bfcV}{\ensuremath{0.40(3)}}

\newcommand{\sgc}{\ensuremath{\sigma_{c}^{\mathrm{cp,g}}}}
\newcommand{\bgc}{\ensuremath{b^{\mathrm{cp,g}}}}

\newcommand{\sgcV}{\ensuremath{0.47(3)}}
\newcommand{\bgcV}{\ensuremath{0.45(5)}}

\newcommand{\stc}{\ensuremath{\sigma_{c}^{\mathrm{cp,3}}}}
\newcommand{\btc}{\ensuremath{b^{\mathrm{cp,3}}}}

\newcommand{\stcV}{\ensuremath{1.18(8)}}
\newcommand{\btcV}{\ensuremath{0.40(4)}}

\newcommand{\stoc}{\ensuremath{\sigma_{c}^{\mathrm{\tau}}}}
\newcommand{\btoc}{\ensuremath{b^{\mathrm{\tau}}}}
\newcommand{\stocV}{\ensuremath{1.06(23)}}

\newcommand{\sgtc}{\ensuremath{\sigma_{c}^{\mathrm{\tau,g}}}}

\newcommand{\sgtcV}{\ensuremath{0.44(8)}}

\begin{document}

    \title{Phase Transitions of Traveling Salesperson Problems
solved with Linear Programming and Cutting Planes}
    \author{Hendrik Schawe}
    \email{hendrik.schawe@uni-oldenburg.de}
    \author{Alexander K. Hartmann}
    \email{a.hartmann@uni-oldenburg.de}
    \affiliation{Institut f\"ur Physik, Universit\"at Oldenburg, 26111 Oldenbug, Germany}
    \date{\today}

    \begin{abstract}
The Traveling Salesperson problem asks for the shortest cyclic tour
visiting a set of cities given their pairwise distances and belongs
to the NP-hard complexity class, which means that \change{with all known
algorithms in the worst
case} instances
are not solveable in polynomial time,
i.e., \change{the problem} is \emph{hard}. Though that does not mean, that
there are not subsets of the problem which are \emph{easy} to solve.
To examine \change{numerically} transitions
from an \emph{easy} to a \emph{hard} phase, \change{a random
ensemble} of cities \change{in the} Euclidean plane
given a parameter $\sigma$, which governs the hardness, is
introduced. \change{Here,}
a linear programming approach \change{together with suitable
cutting planes is applied. Such algorithms operate \emph{outside}
the space of feasible solutions and are often used in practical
application but rarely studied in physics
so far. We observe several transitions.}
 To characterize these
transitions, scaling assumptions from continuous phase transitions
are applied.
    \end{abstract}


    \maketitle


The \emph{Traveling Salesperson Problem} (TSP) \cite{cook2012pursuit}
is to find the shortest
tour through a given set of cities, with known pairwise distances, and
going back to the initial city.
\change{TSP belongs to the class of NP-hard optimization problems
\cite{arora2009computational}, where so far only algorithms with
exponentially growing \emph{worst-case} running time are known. Thus, a
} good tour optimization can not only save money when used for
real world applications,
but also it has a history as a testbed for 
\change{exact \cite{applegate2003implementing,cook2012pursuit}
as well as heuristic
optimization algorithms}, e.g.,
\emph{simulated annealing} \cite{kirkpatrick1983optimization},
\emph{taboo search} \cite{Glover1986Taboo} or
\emph{ant colony algorithms} \cite{dorigo1997ant}.
\change{Also, for the TSP
there exist specific} heuristics
\cite{lin1973effective, dorigo1997ant, edgeElimination}. 
For the Euclidean case (which is still NP-hard \cite{etspNP}),
i.e., the pairwise distances are the Euclidean
distances, a polynomial-time approximation scheme
\cite{arora1998polynomial} \change{is known}. For special corner cases \cite{solvablespecial} even polynomial-time algorithms
exist.

Interestingly, NP-complete problems often show  phase transitions
\cite{hartmann2006phase,mezard2009information,mertens}
\change{where instances are \emph{typically} \emph{easy} to
solve in one region and \emph{typically} \emph{hard} in the other region.}
Some of the classical NP-complete problems \cite{karp1972reducibility}
were examined with respect to phase
transitions with methods of statistical mechanics in Ref.~\cite{martin2001statistical, Biroli, vertexCover2, vertexCover1, krzakala2007gibbs}.
\change{Note that in the statistical mechanics literature usually algorithms
like branch-and-bound \cite{papadimitriou,cover-time2001,cocco2001}, stochastic search
\cite{schneider2006} and message-passing algorithms \cite{mezard2002}
are studied which operate \emph{inside} the space of
feasible configurations. In contrast, for practical applications, algorithms
based on linear programming (LP) dominate because they are very efficient.
These LP-based algorithms operate \emph{outside} the space of
feasible solutions and they should be given more attention in the physics
community. For this reasons
 we study here LP algorithms with respect
to phase transitions for the TSP.}

In Ref.~\cite{Gent} the Euclidean TSP decision
problem on random realizations of cities scattered on the unit square
was under scrutiny and shows a ``transition'' when asking when the tour
length exceeds a certain rescaled threshold. \change{But here the
two ``phases'' are not with respect to basic properties of the instances,
there is no parametrized ensemble. Rather,
the instances are sorted into two classes \change{after}
they are solved, basically reflecting the typical growth of the tour length.}
\change{Instead, here we define a parametrized ensemble of TSP instances.
We study the solvability by a polyonmial-time standard LP approach
together with several types of so-called \emph{cutting-planes}.
We find \emph{several} ``easy--hard''
transitions, similar to one previously found
for the vertex-cover problem \cite{dewenter2012phase,takabe2013}.}

\begin{figure*}[htb]
    \subfigure[$\sigma=0$]{
        \label{sfig:evo0}
        \includegraphics[width=0.23\textwidth]{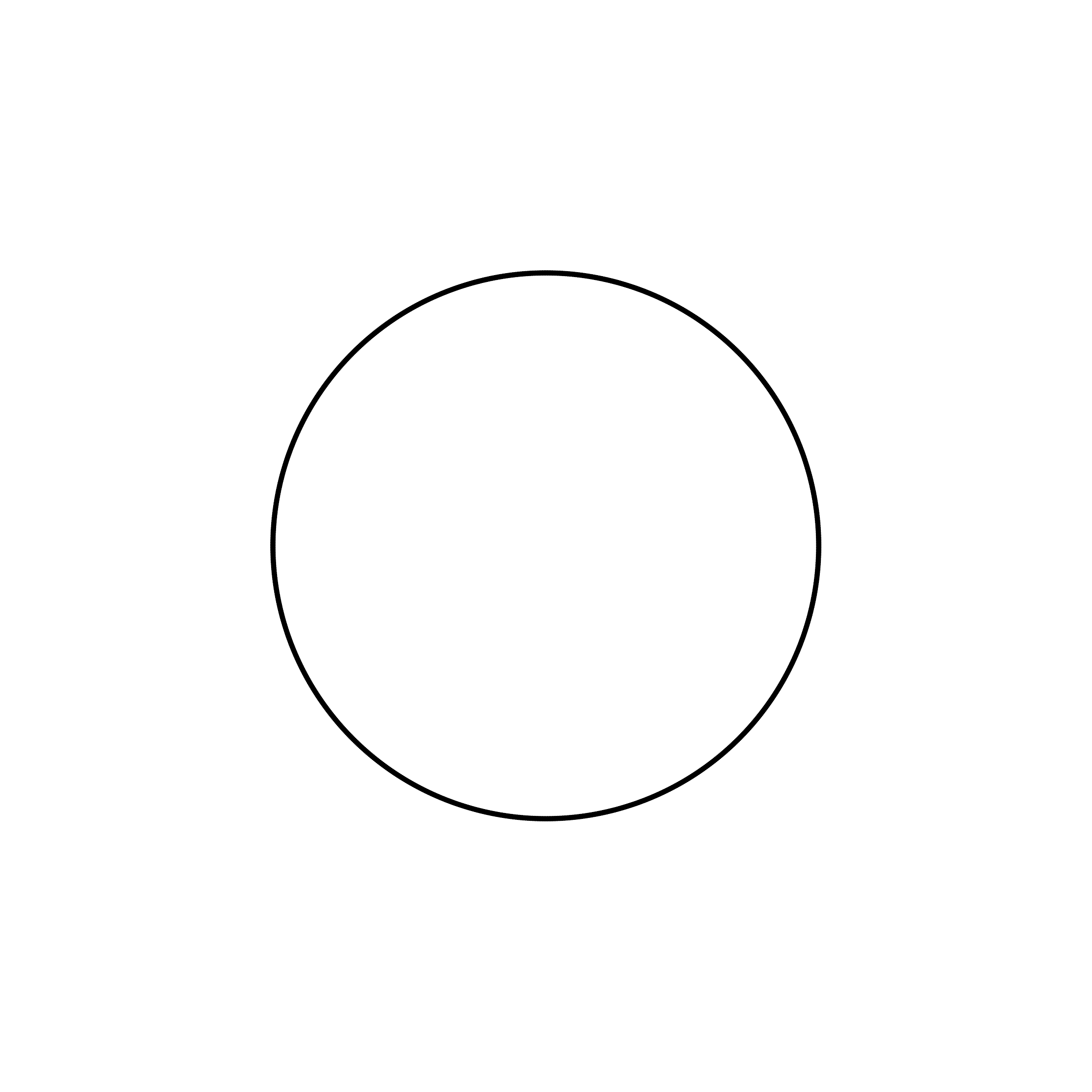}
    }
    \subfigure[$\sigma=20$]{
        \label{sfig:evo20}
        \includegraphics[width=0.23\textwidth]{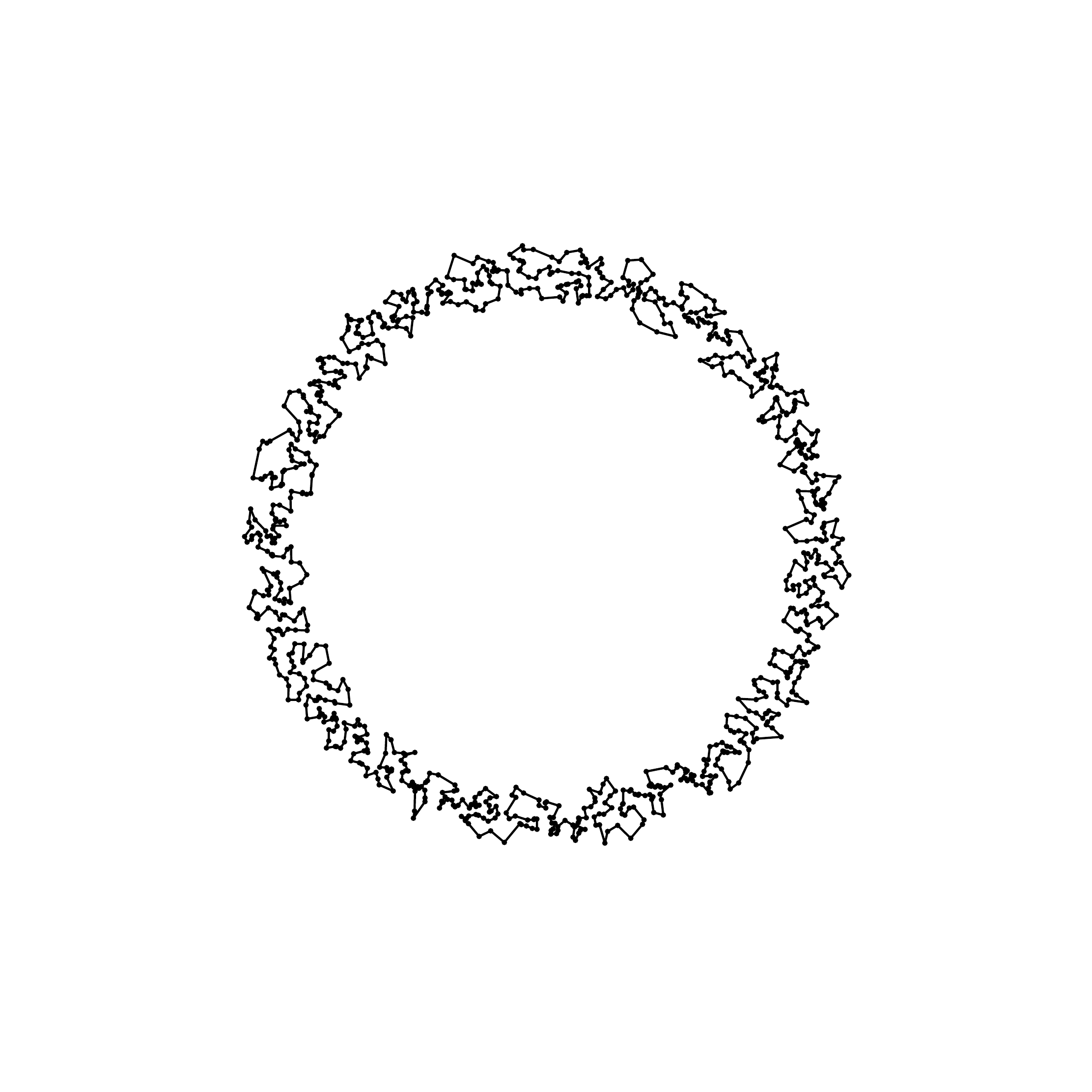}
    }
    \subfigure[$\sigma=80$]{
        \label{sfig:evo80}
        \includegraphics[width=0.23\textwidth]{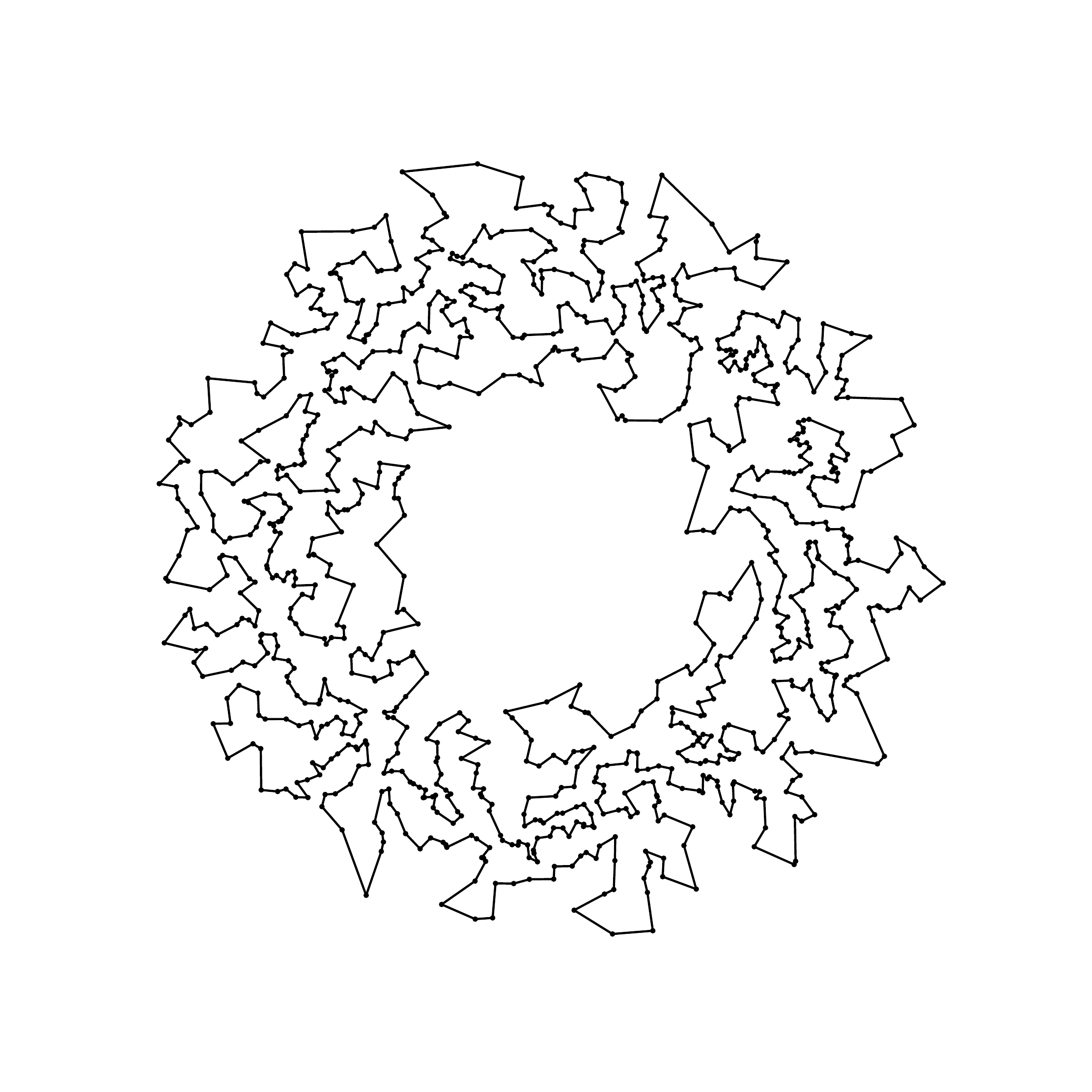}
    }
    \subfigure[$\sigma=160$]{
        \label{sfig:evo160}
        \includegraphics[width=0.23\textwidth]{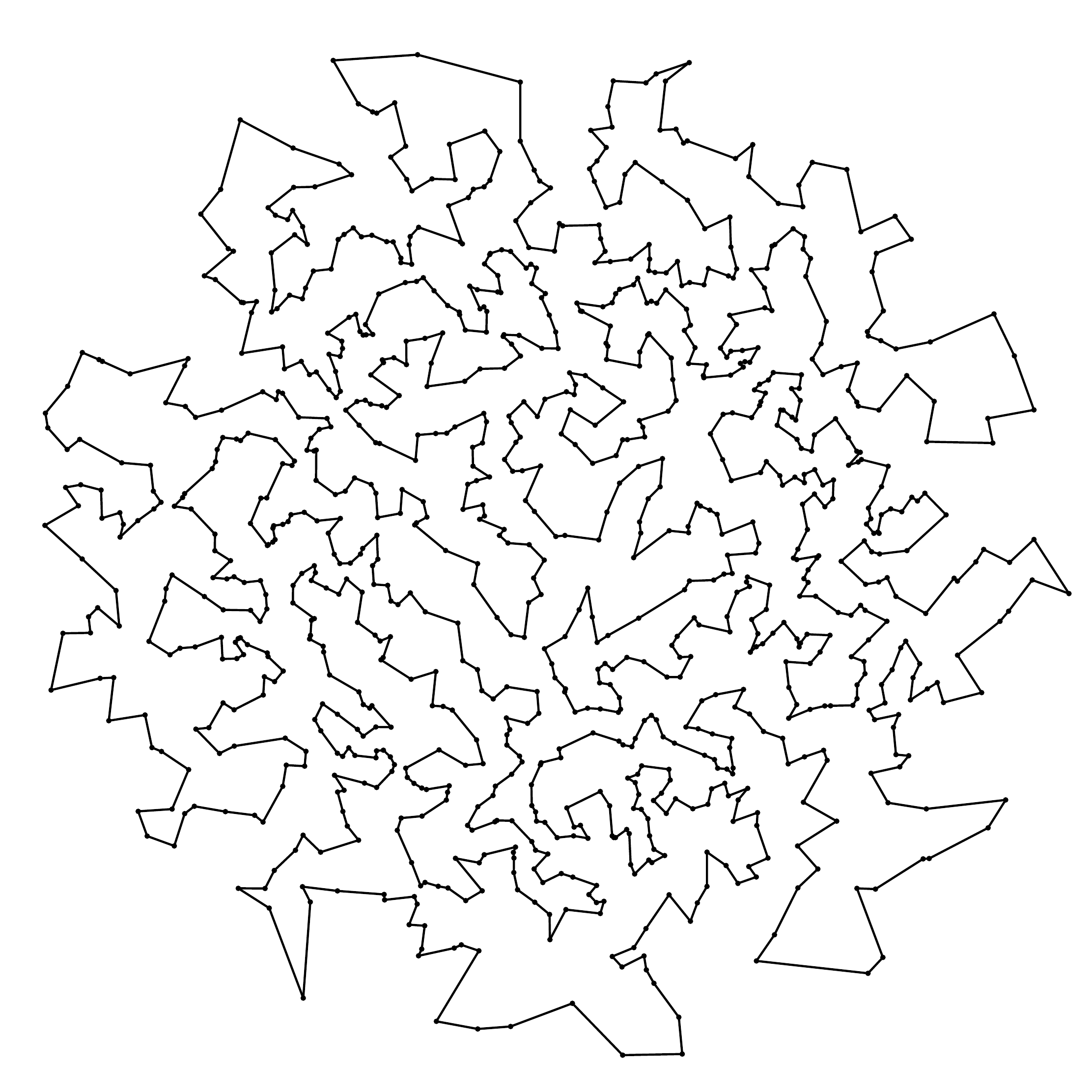}
    }
    \caption{
        Evolution of a $N=1024$ and $R=\frac{N}{2\pi}\approx 160$ system
        with increasing \change{disorder} $\s$. These sample realizations
were solved with \emph{Concorde} \cite{applegate2003implementing}. 
        Obviously the leftmost configuration is \emph{easy} to solve, but the
        other three are probably not.
    }
    \label{fig:evolution1024}
\end{figure*}

The two-dimensional Euclidean TSP is under
\change{scrutiny \cite{percus1996finite, Gent}. Each city from
the set of cities $V$
has coordinates on a plane determining the pairwise distances $c_{ij}$ as their
Euclidean distances, in particular ~$c_{ij} = c_{ji}$.}
\change{We generated each instance of $N$ cities} by random displacement
of cities from a well defined start configuration,
chosen as $N$ cities lying on a circle with a circumference of $2\pi N$,
i.e., the distance between two neighboring cities is approximately $1$.
\change{Note that for the circle}
 even the most simple greedy heuristics, e.g., \emph{nearest neighbor},
finds the optimal tour. Further the circle fulfills the \emph{necklace
condition} \cite{Edelsbrunner1989157} which enables a polynomial-time
solution algorithm and all points are part of the convex hull which
also solves the tour \cite{convexHullTSP}.
For each city the displacement is determined
by two independent random variables from an uniform distribution.
$\phi \in [0, 2\pi)$ is treated
as a displacement angle and $r \in [0, \s]$ as a radius, such that the
new position of a city lies inside of a disk with radius $\s$ around
its initial position. \change{Four sample instances together
with their optimal tours for} $N=1024$ cities are
shown in Fig.~\ref{fig:evolution1024}.

\change{Next, we present our numerical approach}. LPs
can be solved in polynomial-time using the \emph{ellipsoid} algorithm \cite{papadimitriou}.
In this study the \emph{simplex} algorithm \cite{papadimitriou}
implemented by the commercial optimization library CPLEX
is used instead, for its good runtime behavior in the typical case.
If there are constraints which enforce the variables to be integer,
it is called \emph{integer program} (IP) which also belongs to the class
of NP-hard problems.
One can formulate the TSP as an integer program with the objective Eq.~\eqref{eq:objective}
and the constraints Eq.~\eqref{eq:int} to \eqref{eq:sec} \cite{dantzig1954solution}.
\begin{align}
    \label{eq:objective}
    &\text{minimize}     &  \sum_i \sum_{j<i} c_{ij} x_{ij}\\
    \label{eq:int}
    &\text{subject to}   &  x_{ij}                                &\in \{0,1\}\\ 
    \label{eq:inout}
    &                    &  \sum_{j} x_{ij}                       &= 2&            & \forall i \in V \\
    \label{eq:sec}
    &                    &  \sum_{i \in S, j \notin S} x_{ij}     &\ge 2&          & \forall S \varsubsetneq V, S \ne \varnothing
\end{align}
Where the variables $x_{ij}$ are $1$ if $i$ and $j$ are consecutive
in the tour, and $0$ otherwise. The \emph{objective} Eq.~\eqref{eq:objective}
minimizes the tour length. The \emph{integer constraints} Eq.~\eqref
{eq:int} restrict $x_{ij}$ to the integers $0$ and $1$, the \emph
{degree constraints} Eq.~\eqref{eq:inout} ensure that every city has
exactly two neighbors, one for the salesperson to enter one to leave.
And the \emph{subtour elimination constraints} (SEC) Eq.~\eqref
{eq:sec} prevent closed subtours, i.e., loops which visit just a
subset of all cities, by forcing at least two edges to cross the boundaries
of all sets $S \varsubsetneq V, S \ne \varnothing$, which ensures that the salesperson
can enter and leave the set. Hence a closed subtour would violate the constraint
for the set $S$ which contains all cities of the subtour.
Note that there are exponentially many SECs,
because there are exponentially many different subsets $S \subset V$.
To solve this integer program, it is first relaxed to a LP,
i.e., Eq.~\eqref{eq:int} is replaced by $x_{ij} \in [0,1]$.
The solution of this LP relaxation
will always have a better or equal tour length than the solution of the
TSP, but may not always be a valid tour, i.e., may have fractional $x_{ij}$.
Though, if the solution of a LP relaxation is integer, it is guaranteed
to be the optimal tour.

Because there are exponentially many SECs, they will not be enforced
in the beginning, instead SECs \change{will be added if}
violated by the current LP solution, \change{and the resulting LP is
solved again. The violated SECs}
can be found by a \emph{global minimum cut}, e.g., with the St{\"o}r-Wagner
algorithm in polynomial-time \cite{stoer1997simple}.
This is iterated until no violated SEC exists anymore.

A measure of hardness \change{of an instance for a given
LP algorithm is as follows:}
if the LP relaxation
\change{results in all variables being integer}, i.e.,
if the instance can be solved in polynomial time
\cite{grotschel1981ellipsoid,grotschel1993geometric,Conforti2014},
it is therefore \emph{easy}.
Also we will look at the \emph{degree LP relaxation} where the SECs are
removed and only the degree constraints \eqref{eq:inout} and the bounds are enforced.
\change{Here, we also find instances which are solved by this
simpler algorithm. Thus, they
can be considered even \emph{easier}.}

Note that this algorithm can easily be extended to
find always the optimal solution, by a branch-and-cut
search \cite{applegate2003implementing} \change{at the cost of
a worst-case exponential running time. Nevertheless, here we are mostly
interested in the algorithm-dependent hardness of an instance, not
necessarily in always finding a solution.}
%
%
%
The focus on the \change{solvability} by LP methods allows
reasonable big instances of up to $N = 1448$
cities at $80$ different $\s \in [0,60]$ and $5000$ samples each.
All errorbars are obtained via bootstrap resampling \cite{efron1979,practicalGuide,young2015}
if not noted otherwise.

\begin{figure}[htb]
    \includegraphics[scale=1]{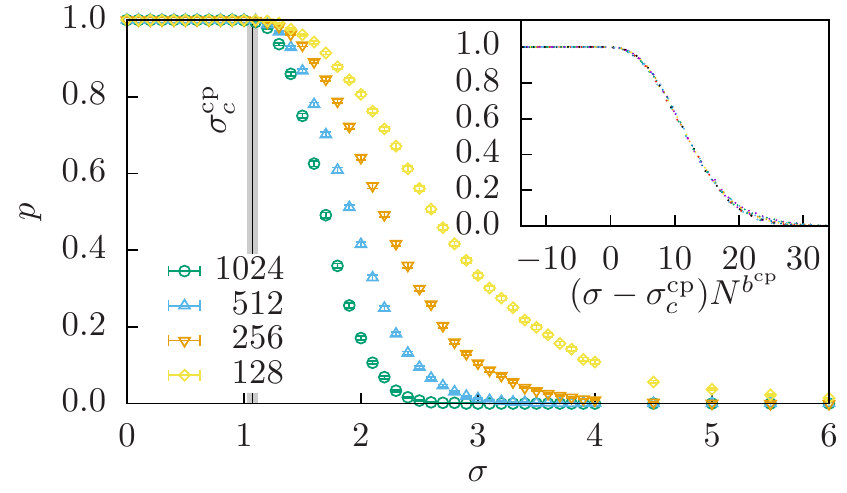}
    \caption[Probability $p$ that the LP relaxation is Integer]
    {
        (color online) 
        Probability $p$ that the LP relaxation is integer, i.e., the
        solution can be obtained by LP, as a function of the displacement
        parameter $\s$.
        The inset shows the same $p$ for $N \ge 256$ plotted with a rescaled $\s$-axis
        with $\sac = \sacV$ and $\bac = \bacV$ obtained by Fig.~\ref{fig:pLPpeaks}.
        Different symbols and errorbars are omitted for clarity.
    }
    \label{fig:pLP}
\end{figure}
The probability $p$ to find the \change{true integer}
solution using the LP relaxation
is plotted in Fig.~\ref{fig:pLP}. For small disorder,
$p$ is constant at $p=1$ and falls with increasing $\s$ to $p=0$.
With increasing system size $N$ the \change{curves become} steeper.
This pattern is typical for a phase transition. \change{Therefore,
the results indicate}
a phase transition from an \emph{easy} \change{phase, where
the instances are typically} solvable by
polynomial-time linear programming techniques, to a \emph{hard} phase.
\begin{figure}[thbp]
    \includegraphics[scale=1]{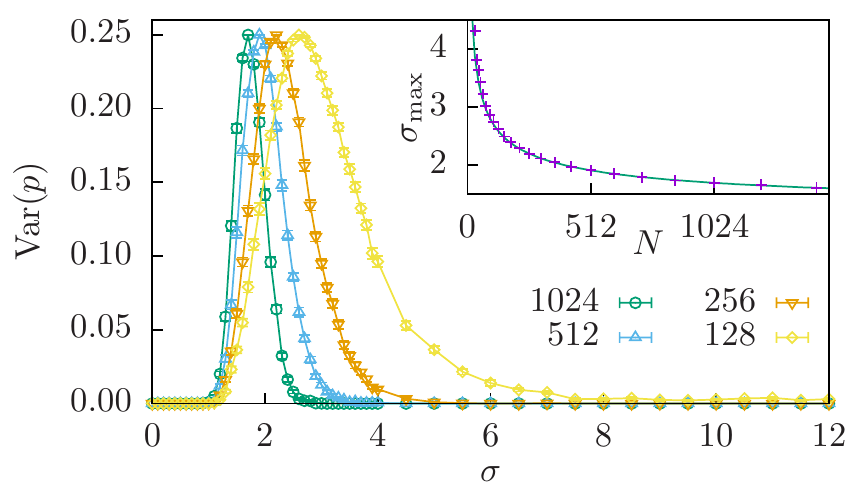}
    \caption
    {
        (color online)
        Variance of the solution probability $\Var(p)$ as a function
        of the disorder $\s$.
        The inset shows the position of maximal variance of the solution
        probability $p$ over the number of cities $N$.
        The positions are obtained by second order polynomials
        fitted to the 5 data points next to the peak.
        The power-law $\s = aN^{-\bac} + \sac$ is fitted to
        the peak positions for $N \ge 256$ to minimize the effects of
        corrections to scaling, yielding $\sac = \sacV$ and $\bac = \bacV$.
    }
    \label{fig:pLPpeaks}
\end{figure}
\change{Next, we determined}
the transition point $\sac$ \change{in the limit $N \to \infty$} and the
exponent $\bac$, \change{governing the finite-size scaling behavior
\cite{goldenfeld1992} near the
transition point, corresponding to the correlation-length
exponent for physical systems. For this purpose we fitted parabolas}
 to the
variance of $p$ in
vincinity of the maximum, \change{see} Fig.~\ref{fig:pLPpeaks}.
For second-order phase transitions the peak
positions are \change{expected to follow}
$\s = \sac + aN^{-\bac}$, \change{which holds well for our data}
 as depicted in the
inset of Fig.~\ref{fig:pLPpeaks}.

\change{According to finite-size scaling,
rescaling the $\sigma$ \change{axis} according to $(\s-\sac)N^{\bac}$
should yield a collapse of the data onto one curve \cite{Heermann}}
for
big values of $N$ in vicinity of the critical point.
\change{This is true for our data as visible in the}
inset of Fig.~\ref{fig:pLP}.
confirming
the values of $\sac$ and $\bac$.

\begin{figure}[htb]
    \includegraphics[scale=1]{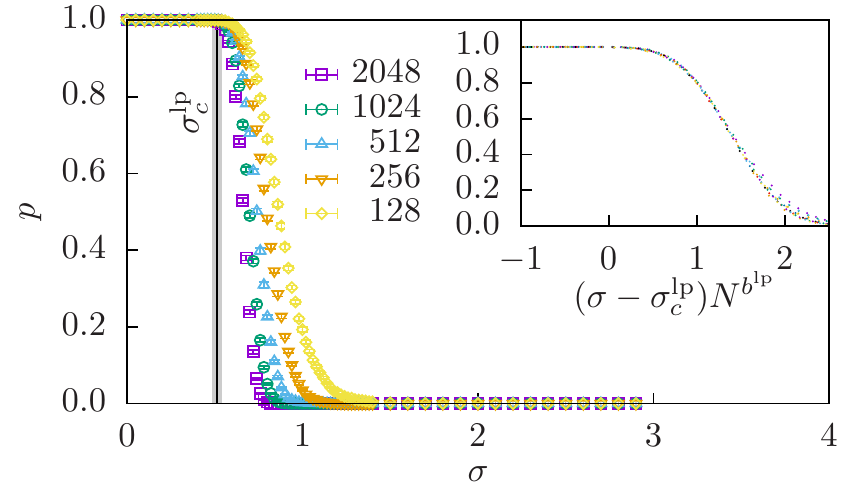}
    \caption
    {
        (color online)
        Probability $p$ that the degree LP relaxation is integer.
        The inset shows the collapse for $N \ge 256$, with
        $\sdc = \sdcV$ and $\bdc = \bdcV$ obtained from the same type of
        analysis as in Fig.~\ref{fig:pLPpeaks} for $N \ge 512$.
    }
    \label{fig:pLPLP}
\end{figure}

\change{To identify a region of
even \emph{easier} instances,
we studied also the LP with applying only the degree contraints
\eqref{eq:inout}, see Fig.~\ref{fig:pLPLP}. We found a second easy--hard
transition with} $\sdc = \sdcV$ and $\bdc =
\bdcV$.

A further class of \change{cutting-plane}
inequalities for the TSP are blossom inequalities \cite{padberg1982odd}
which originate from the two-matching LP \cite{edmonds1965maximum}.
A subset, which
is easy to separate using heuristics, are \emph{fast blossoms}
\cite{applegate2003implementing},
available in \emph{Concorde} \cite{applegate2003implementing}.
Doing the same analysis as above
 \change{revealed a third transition (not shown)}
 at $\sfc = \sfcV$ with $\bfc = \bfcV$.

\change{Next, we want to find out whether the easy--hard transitions
are accompanied by changes of suitably defined structural order parameters.}
For up to $N=180$ the optimal tours for all studied samples 
were obtained by a branch-and-cut
procedure, available in \emph{CPLEX},
 to examine structural properties of the solutions.
With increasing \change{value of} $\s$
 optimized tours appear to be
 more ``meandering'' as shown in Fig.~\ref{fig:evolution1024}.
As a measure of
this ``meandering'', we used the \emph{tortuosity}, as defined in 
Ref.\ \cite{tortuosity}, where it was used to evaluate images of blood vessels
in the retina to detect vascular diseases.
To calculate $\tau$, the tour is segmented into $n$ segments, \change{such that
each segment has the same curvature sign and is of maximal length.}
Let the \emph{arc length} $L_i$ be the length of the segment $i$
along the tour and
let the \emph{chord length} $S_i$ be the direct distance between the first
and last city of the segment $i$ and $L$ the total length of the tour.
Then the tortuosity is defined as
\begin{align}
    \tau = \frac{n-1}{L} \sum_{i=1}^{n} \brac{\frac{L_i}{S_i}-1}.
    \label{eq:tortuosity}
\end{align}

\begin{figure}[htbp]
    \includegraphics[scale=1]{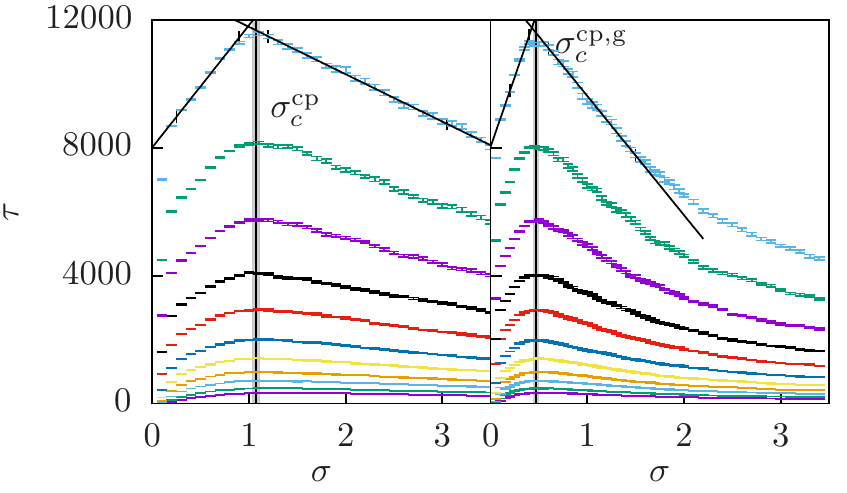}
    \caption
    {
        (color online)
        The tortuosity $\tau$ of optimal tours as a function of the
        displacement parameter $\s$ for different system sizes $N \le 180$
        peaks at $\stoc$, which coincides with $\sac$ determined
        by the fit in Fig.~\ref{fig:pLPpeaks} (gray bar $=$ mean $+$ error).
        On the right side the same is done with data from a slightly
        modified model, explained later in the text.
    }
    \label{fig:tortuosity}
\end{figure}
When plotting $\tau$ as a function of $\s$ in Fig.~\ref{fig:tortuosity},
it shows peaks near $\sac$. As a very rough estimate of the
position of this peak, straight lines are fitted to $\tau$ at $N=180$
left and right of the peak and their intersection is interpreted as
an estimate of the peak positions, with errors obtained by error propagation.
This is shown for $N = 180$ in Fig.~\ref{fig:tortuosity} and done for all
sizes $N \ge 64$.  \change{Via a power-law fit to $\st =   \stoc+aN^{-c}$,}
we estimated an asymptotic $\stoc = \stocV$, which is consistent
with the estimate $\sac$ from Fig.~\ref{fig:pLPpeaks},
 Unfortunately the fit is not good enough to give a
meaningful estimate of the more susceptible corresponding exponent $\btoc$.

Comparing the solution tour $x_{ij}$ to the
\change{circular shaped}
optimal tour at $\s=0$, it is expected that they are similar at very small
\change{disorder} $\s$.
A way to measure this similarity is to look at the number of edges occurring
in the one tour but not in the other, i.e.,
the Hamming distance \cite{hamming1950error}.
\begin{figure}[htbp]
    \centering
    \includegraphics[scale=1]{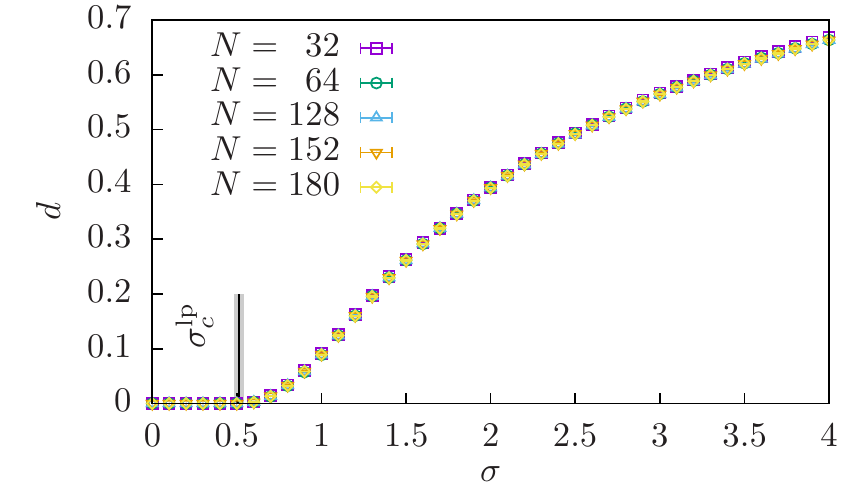}
    \caption
    {
        (color online)
        Difference of the optimal tour to the initial circle as a function
        of $\s$.
    }
    \label{fig:diff}
\end{figure}
The tour difference $d$ shown in Fig.~\ref{fig:diff} is the Hamming distance
normalized by $2N$, such that two tours with no common edges would result
in $d=1$ while two tours visiting the cities in the same sequence would result
in $d=0$. This observable seems to be \change{roughly}
independent of $N$. Fig.~\ref{fig:diff}
suggests that \change{the easy--hard transition observed when using
 the degree LP relaxation alone corresponds to the structural change
observed by studying the hamming distance $d$.}

Unfortunately, \change{we were not able to indentify so far an observable}
 which corresponds to the phase transition
\change{occurring when using} the fast blossoms.


\change{We performed the same analysis for
a different random ensemble, where the cities are displaced by}
 $\Delta x$ and $\Delta y$
from a Gaussian distribution $G(0, \s)$ \change{for each} direction.
As expected for continuous phase transitions, \change{we obtained
(not shown)}
the same critical exponent $\bgc = \bgcV$ within errorbars,
which hints that this model \change{exhibits universality with
respect to the type of disorder. The resulting values are}
 shown in Tab.~\ref{tab:values}.
Also for the case of cities displaced spherical
by $\phi$, $\theta$ and $r$ from uniform
distributions in three dimensions it shows the same critical exponent
$\btc = \btcV$. Note that unlike many other models
(e.g.~Ising ferromagnet or percolation)
the different dimension does not lead to a different exponent.

\begin{table}[htb]
    \begin{ruledtabular}
        \begin{tabular}{l r@{}l r@{}l}
             & \multicolumn{2}{c}{\(\sC\)} & \multicolumn{2}{c}{\(\bC\)}\\
            \colrule
            with SEC        & $\sac  =$&$ \sacV$  & $\bac =$&$ \bacV$ \\
                            & $\stoc =$&$ \stocV$ & & --\\
                            & $\sgc  =$&$ \sgcV$  & $\bgc =$&$ \bgcV$ \\
                            & $\sgtc =$&$ \sgtcV$ & &--\\
                            & $\stc  =$&$ \stcV$  & $\btc =$&$ \btcV$ \\
            \colrule
            only degree     & $\sdc  =$&$ \sdcV$  & $\bdc =$&$ \bdcV$ \\
            \colrule
            fast blossoms   & $\sfc  =$&$ \sfcV$  & $\bfc =$&$ \bfcV$ \\
        \end{tabular}
    \end{ruledtabular}
    \caption{
        Values of critical points and exponents grouped by different
        types of transitions.
    }
    \label{tab:values}
\end{table}


We have shown that for this random ensemble governed by the parameter $\sigma$
there exist \change{various easy--hard
phase transitions. This indicates a rich behavior of the ensemble
with respect to the typical computational hardness. Furthermore,
at least for two cases we found that the transitions can be correlated
with measurable changes of the solution structure, namely Hamming distance
to the circle solution and tortuosity, respectively. The
transitions} can be characterized by
critical exponents $b$.
 \change{Within the statistical accuracy of our data,
the critical exponents for the different easy--hard transitions are
compatible within two sigma.}

An interesting question for further study would be finding an answer to
why the tortuosity $\tau$ peaks at $\sigma_c$, where the TSP becomes
not solvable using LP and SEC. Unfortunately $\tau$ is
quite complex to measure.
 Therefore the search for a simpler observable showing the
transition would be of equal interest.

Besides the blossom inequalities, there are more
complicated
inequalities valid for the TSP establishing facets on the polytope, which
can be implemented as cutting planes and partly already be separated in
polynomial-time \cite{fleischer2006polynomial}.
It would be interesting if those would establish a further phase transition
at higher $\sigma$ and if the critical exponent $b$ stays the same.

\change{In general, LP-based algorithms are used a lot in practice and it
would be of great interest to study suitable ensembles of other
NP-hard
optimization problems with respect to easy--hard transitions. Furthermore,
a statistical-mechanics analysis of the performance of LP-based
algorithms, like done in the past for branch-and-bound algorithms
\cite{cover-time2001,cocco2001},
would yield more insight into the sources
of computational hardness.}

    \section*{Acknowledgments}

The simulations were performed at the HPC Cluster HERO, located at
the University of Oldenburg (Germany) and funded by the DFG through
its Major Research Instrumentation Programme (INST 184/108-1 FUGG)
and the Ministry of Science and Culture (MWK) of the Lower Saxony
State.
    \bibliography{lit}

\end{document}